\documentclass[a4paper,fleqn,usenatbib,useAMS]{mnras}

\usepackage{graphicx}

\usepackage[T1]{fontenc}
\usepackage{ae,aecompl}

\title[Torques:counter-rotation and ring galaxies]{Torques and angular momentum: 
Counter-rotation in galaxies and ring galaxies}

\author[N. G. Kantharia]{N. G. Kantharia$^{1}$\thanks{Contact e-mail: \href{mailto:nkprasadnetra@gmail.com}{nkprasadnetra@gmail.com}}\\
$^1$ National Centre for Radio Astrophysics, TIFR, Post Bag 3, Ganeshkhind, Pune-411007, India}

\date{Last updated; in original form }

\pubyear{2016}

\begin{document}
\maketitle

\begin{abstract}

We present an alternate origin scenario to explain the observed phenomena of (1) 
counter-rotation between different galaxy components and (2) the formation of ring 
galaxies.  We suggest that these are direct consequences of the galaxy being acted 
upon by a torque which causes a change in its primordial spin angular momentum  
first observed as changes in the gas kinematics or distribution.  We suggest that 
this torque is exerted by the gravitational force between nearby galaxies. This 
origin requires the presence of at least one companion galaxy in the vicinity - we 
find a companion galaxy within 750 kpc for 51/57 counter-rotating galaxies and 
literature indicates that all ring galaxies have a companion galaxy thus giving 
observable credence to this origin.  Moreover in these 51 galaxies, we find a 
kinematic offset between stellar and gas heliocentric velocities $>50$ 
kms$^{-1}$ for several galaxies if the separation between the galaxies $<$ 100 kpc.
This, we suggest, indicates a change in the orbital angular momentum of
the torqued galaxies. 
A major difference between the torque origin suggested here and the existing model of 
gas accretion/galaxy collision, generally used to explain the above two phenomena, 
is that the torque acts on the matter of the same galaxy whereas in the latter 
case gas, with different properties, is brought in from outside.  An important 
implication of our study is that mutual gravity torques acting on the galaxy can 
explain the formation of warps, polar ring galaxies and lenticular galaxies.  We 
conclude that mutual gravity torques 
play an important role in the dynamical evolution 
of galaxies and that they naturally explain several galaxy observables.

\end{abstract}


\section{Introduction}

\subsection{Counter-rotation in galaxies}
Since their discovery in the 1980s \citep{1984Msngr..37...17B,1986ApJ...305..136C,1987ApJ...318..531G}, 
several galaxies showing counter-rotation between the main stellar disk and the gaseous component have 
been identified.   Two main models have been put forward to
explain this phenomenon: (1) external gas accretion of opposite spin either through continuous
accretion or a merger of a counter-rotating galaxy as suggested in the first cases of
identified counter-rotation and (2) internal bar which can support 
different spins \citep{1994ApJ...420L..67E}.  
However we find that there are a few key observational signatures of counter-rotating
galaxies which these models fail to satisfactorily explain which prompted us to
revisit the origin of counter-rotating material in galaxies.
 
The main observables of a counter-rotating galaxy can be summarised as:
(1) undisturbed optical disk morphology  (2) gas/stars
are rotating in the opposite sense compared to the main stellar disk
(3) generally gas-poor (4) the counter-rotating
population of stars is generally younger than the main stellar disk (5) the
counter-rotating population of stars, are sometimes deduced to be of slightly lower metallicity
compared to the main stellar disk.   We believe that 
observable (1) which seems to be universal is a severe shortcoming of the 
merger model which should affect the optical morphology.  We find that observable (3) is a shortcoming
of the continous gas accretion model since in this case the counter-rotating
galaxy should be gas-rich which is generally not the case.  
However we also note that some of these observed characteristics need to be confirmed by
new observations in particular, the metallicity differences.  Considering many early type galaxies
are also now known to harbour gas,  the gas-poor property of counter-rotating galaxies 
also needs to be revisited. 

Simulations support gas accretion either through continuous infall or a dwarf merger 
\citep[e.g. NGC 4138;][]{1996AJ....112..438J, 1997ApJ...479..702T}.  In the merger case, the 
undisturbed optical morphology is explained by letting $> 1$ Gyr elapse between the merger and now 
and hence predicts that the counter-rotating stellar population will have contribution from 
stars of age $\sim 1$ Gyr.  If there is continuous counter-rotating gas accreted over Gyrs
\citep[e.g.][]{1997ApJ...479..702T}, then 
we need to understand the process of where the gas is coming from 
and why it is being accreted on retrograde orbits. 
Moreover if there was continuous gas accretion the counter-rotating galaxies should be gas-rich
systems.   If accreted gas is from a nearby galaxy then the timescales for the gas to be
tidally braked and stripped from the other galaxy + travel to the counter-rotating galaxy +
settling down has to be taken into account.  It would also indicate that the star formation 
triggered in the tidal features external to the counter-rotating galaxy will be older than the stars 
inside the counter-rotating material of the galaxy.   All this has to be verified using
observational data. 

Thus, while we agree that retrograde gas accretion can be one of the causes, with the
aforementioned concerns, we are not convinced that it is the most viable explanation for the observed
counter-rotation.  We explore other known physical explanations for counter-rotation. 
In particular, we short-list an external torque acting on the galaxy
as being an important physical process that
can explain counter-rotation.  The importance of such torques has generally been
recognised as one of the primordial causes of rotation in galaxies \citep{1969ApJ...155..393P,
1978ApJ...223..426G}. 

\subsection{Ring Galaxies}
Literature refers to two types of ring galaxies: resonance ring galaxies and collisional
ring galaxies.  The resonance ring galaxies are understood with an origin internal to the 
galaxy which is generally found to host a bar.  The collisional ring galaxies,
on the other hand, are generally explained by a head-on collision of a large disk galaxy 
with a smaller galaxy \citep[e.g.][]{1976ApJ...209..382L}.  However, we note that Vorontsov-Velyaminov, 
who identified several ring galaxies \citep{1960SvA.....4..365V}, after Zwicky's 
identification of the Cartwheel galaxy in 1941, had favoured an origin
of ring galaxies as 'not catastrophic but calm and without external influence'  
\citep{1976SvAL....2..204V}.  We agree with the first part of his statement about the
evolution being calm but as we show it is likely to have had an external trigger.  

Resonance ring galaxies show the presence of several rings inside the galaxy and lack
of nearby companions, while
collisional ring galaxies show the presence of a ring external to the optical galaxy and 
presence of nearby companions.  In this paper we are concerned with the collisional ring galaxies.

Main observables of collisional ring galaxies can be summarised to be: 
(1) a ring of material located outside the optical galaxy 
observable in atomic hydrogen HI, H$\alpha$ and optical photometric bands 
(2) blue colours of the ring inferred to be due to a
younger stellar population  (3) expanding ring  (4) rotating ring with rotation typical
of the parent galaxy (5) a closeby companion 
(6) Low column density HI and/or
faint photometric emission from the region between the galaxy and the ring.  

While the head-on collision might be a possible origin for these class of galaxies, 
we find it difficult to reconcile observables (4) and (5) with that
model.  Additionally we note that the ring galaxies are generally gas-poor and do not appear disturbed.    
Thus, we search for an alternate formation scenario and as in the case of
counter-rotation in galaxies, we again shortlist external torques acting on the galaxy
as being able to explain the formation of these galaxies.  

In the following sections we explain how torques can give rise to several
observables and then use existing observational data from 
literature to show that the mutual gravity torque origin can explain observations.

\section{Torques on galaxies} 
We note that one of the possible origin of disk galaxy
rotation has been shown to be torques due to nearby neighbours 
\citep{1969ApJ...155..393P,1978ApJ...223..426G} and we find no compelling reasons to
believe that they are not important in the dynamical evolution of galaxies in the present epoch.  
Considering that more than 60\% of galaxies reside
in groups, the effect of torques on galaxies should actually be ubiquitous and likely responsible
for several observable characteristics.  Even 
timescales do not constitute a problem since these torques can be effective over timescales ranging from
100 million years to few Gyrs.   
In the case of galaxy formation, these changes are believed to have occurred over a short 
timescale \citep[$\sim 10^8$ yrs;][]{1969ApJ...155..393P,1967ApJ...147..868P}.  These effects
will precede tidal stripping of gas.  

\begin{figure*}
\includegraphics[width=12cm]{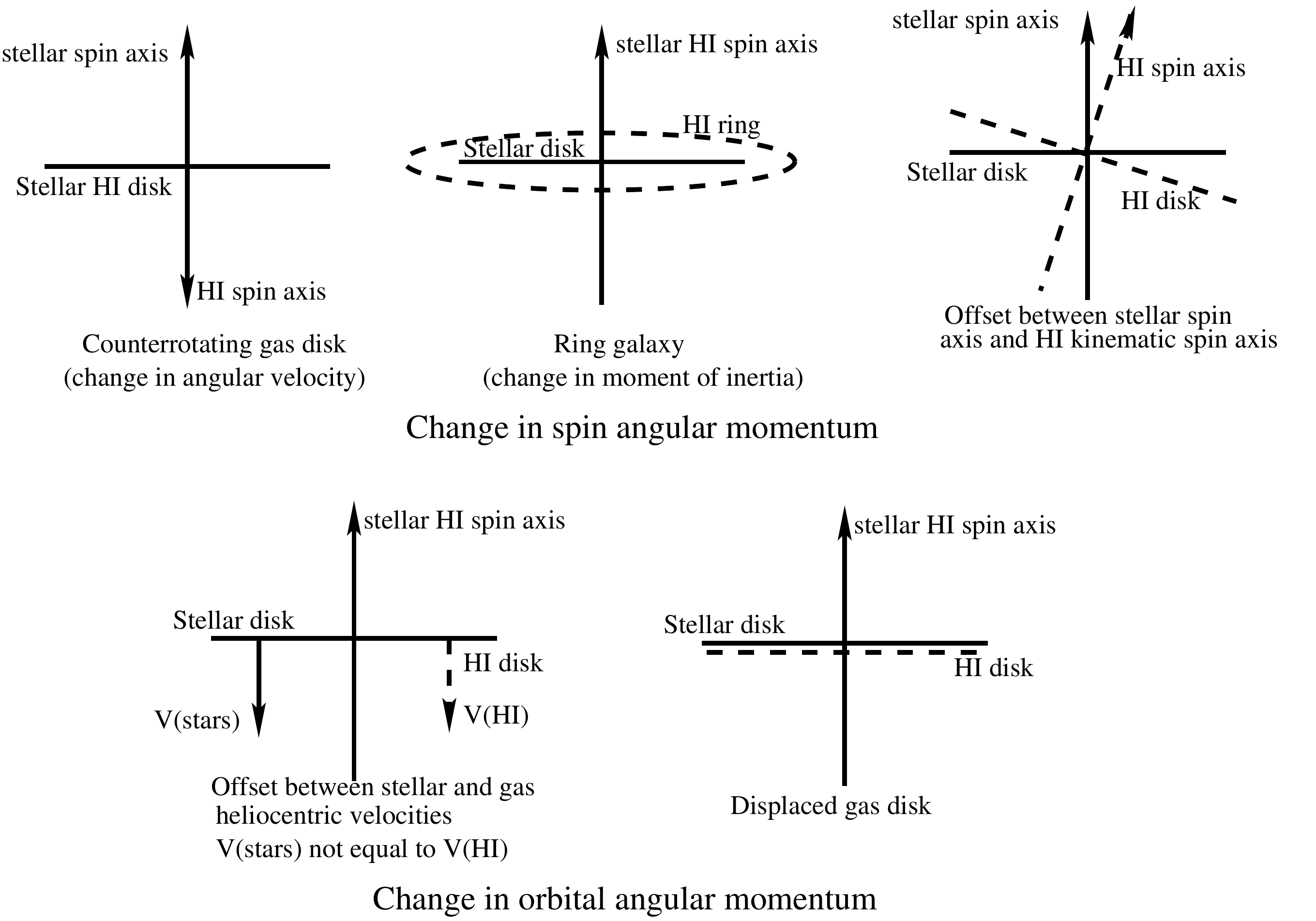}
\caption{Summary of how a torque can modify the spin and orbital angular momentum
in galaxies and the observable features associated with those as discussed in the paper.  
The terms HI and gas refer to the gas disk.
The top row shows the effect of torque on the spin angular momentum of the galaxy and
explains (from left to right) the counter-rotation in galaxies, formation of ring galaxies
and possibly warps and polar ring galaxies.  The lower row
of schematics explains the observables which indicate a change in the orbital angular
momentum of the galaxy.  We note that such features are regularly observed in galaxies. } 
\label{fig1}
\end{figure*}

We, thus, explore the possible outcome of a torque acting on 
a galaxy (G$_1$) due to the passage of another galaxy (G$_2$).  This torque $\tau$ acting on 
G$_1$ can lead to a change in its angular momentum $L$ and conservation of angular momentum 
of the system indicates that it can be balanced by some change in G$_2$.
Since angular momentum of any galaxy
is the combination of spin and orbital angular momentum,  the torque can mediate a transfer 
between the spin and orbital angular momenta or between any one of them.   
The differentially rotating disk galaxies can be 
approximated by annular rings around the centre of the galaxy 
and $\omega=2\pi / T$ will be the angular velocity of material (T is the rotation period) 
in G$_1$ at distance $r$ and $ I=mr^2$ will be its moment of inertia. 
Thus the torque acting on matter in an annular ring in a galaxy G$_1$ will be:
\begin{equation}
\tau = \frac{dL}{dt} = \frac{d(I\omega)}{dt} \\ = \omega \frac{dI}{dt} + I \frac{d\omega}{dt}
\label{eqn1}
\end{equation} 
A similar expression will apply to G$_2$.  The angular momentum will be transferred to the matter
at different radial distances rotating with different angular velocities.  
When required, we use the simplifying assumption that
only two galaxies are involved in the interaction so that the angular momentum conservation
requires changes in the angular momentum of only those two galaxies.  However we note that
in some cases, the torque might be a result of multiple galaxies. 

Equation \ref{eqn1} has two terms and shows that the torque can 
facilitate angular momentum exchange between two nearby galaxies through changes
in two independent physical properties of the galaxy namely moment of inertia $I$ and angular velocity
$\omega$.  Since no change in mass of the galaxy is expected, we use the radial distribution of
gas as the main galaxy property which changes the moment of inertia. 
Thus the torque can lead to change in any of these.  To elaborate, if the torque leads to: 

(1) A change in the spin angular momentum then one of the
following can happen.  If we assume that there is no change in the moment of inertia of
G$_1$ then the torque can modify the spin angular velocity and the kinematics of the gas
will first change.  This change can either speed up the rotation or slow it down leading to a kinematic
decoupling between the gas and stellar disk in G$_1$. 
In the extreme case, it can slow down the 
rotation considerably and even establish a retrograde motion in G$_1$. 
While we believe there might be favourable configurations and kinematics of the pair of galaxies which will
cause the torque to lead to counter-rotation
(e.g. Figure \ref{fig2}), in this paper we focus on the observational evidence 
in support of a torque due to the gravitational force between two galaxies.
On the other hand, if there is no change in $\omega$ in G$_1$ due to the torque,
then there could be a change in the radial distribution of the matter.  The matter in G$_1$ 
can fall towards the centre or move outwards under the effect of a decrease or increase in its angular momentum. 
This could lead to formation of circumnuclear rings and ring galaxies. 
A simple scehmatic showing the observables for a spin angular momentum change are
shown in the top row of Figure \ref{fig1}.

(2) A change in the orbital angular momentum due to the torque 
will also lead to two distinct observational signatures.
If there is a change in the kinematics with no change in moment of inertia - then this could be viewed
as a change in the orbital velocity of the gas disk which would be one of the first
manifestations.  An offset between the heliocentric velocities
measured for the stellar disk and the gas disk would then result.  If there is no detectable 
change in the kinematics,
then the change in its moment of inertia can be mediated by redistributing the gas disk. 
One possible observable would be a displacement between the optical disk and the gas disk seen in HI
so that the orbital separation between the two galaxies changes. 
A simple scehmatic showing the observables for an orbital angular momentum change are
shown in the bottom row of Figure \ref{fig1}.

The gravitational torque will act on all the matter in the galaxy but will result in the first
observable signatures on the gaseous component which will induce differences in the nature of the
gaseous and stellar components.  
One way to understand this is that efficient angular momentum
transfer is possible in the gas disk which is a collisional deformable system 
as compared to the stellar disk
which is collisionless.  Moreover since the mass of the gaseous disk is $\le 1\%$ of the stellar disk,
the effect of the gravitational torque can be manifest on a significant fraction of the gas disk. 
In this paper, we mainly discuss the observable changes in the gaseous component due to the mutual gravity torque.

A mutual gravity torque can lead to a change in the kinematics of the gas in one galaxy and
change in moment of inertia in the other galaxy or can even lead to change in both components 
in both the galaxies.  We suggest that the well-understood physical process of
torques when applied to galaxies, provide a consistent and simple formation scenario
for (1) the observed kinematic decoupling including counter-rotation between, the gas disk 
and the stellar disk and
(2) the formation of ring galaxies. Torques also lead to other observed characteristics (e.g. Figure
\ref{fig1}), but in this paper we test our hypothesis with respect to these two peculiar 
features observed in galaxies.

\subsection{Quantifying torque exerted by gravitational force between two galaxies}
\begin{figure}
\includegraphics[width=7cm]{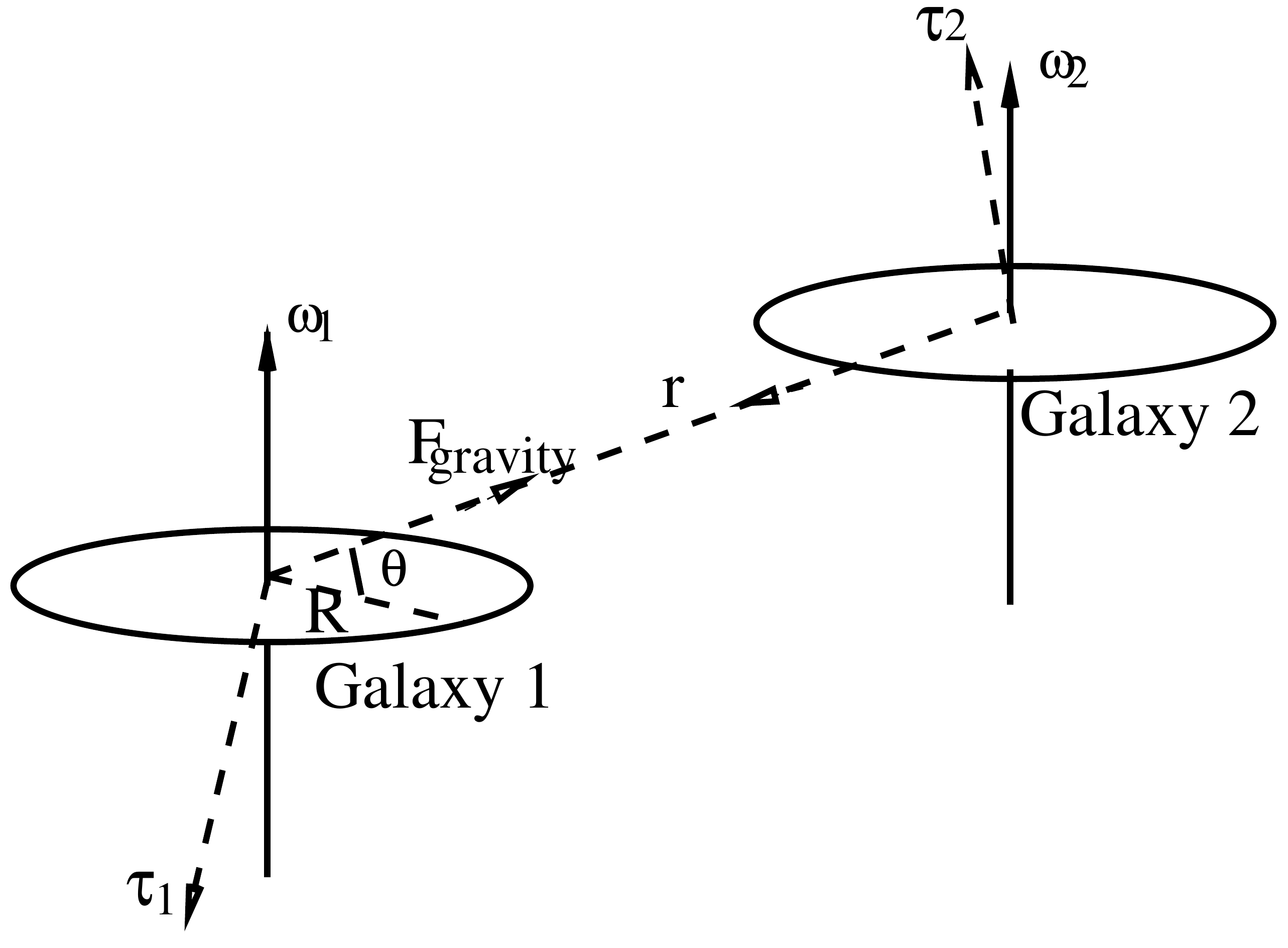}
\caption{A simplified schematic to demonstrate the proposed explanation for
counter-rotation and ring galaxies.   The figure shows the torque $\tau$ that will act on the two galaxies
due to the mutual gravitational force  F$_{gravity}$ which is estimated assuming mass is
concentrated at the centre of mass.  This force exerts a torque on matter in the entire galaxy 
and changes the galaxy's angular momentum.  $\theta$ is the angle made
by F$_{gravity}$ with the galaxy disk. } 
\label{fig2}
\end{figure}

To quantify the effectiveness of a gravitational force and resulting torque on a galaxy, we
use the simple expression of torque which includes force.  This is also useful
in visualising the directions of the force between two galaxies and deducing the 
direction of the torque using the right hand rule (see Figure \ref{fig2}).  

The gravitational force between two galaxies of masses
M$_1$, M$_2$ and separated by a distance r is:

\begin{equation}
F_{gravity} = \frac{G M_1 M_2}{r^2}
\end{equation}

The centrifugal force acting on a cloud of mass m with a rotation velocity
of V located at a distance R from the centre of a galaxy G1 is:

\begin{equation}
F_{centrifugal} = \frac{m V^2}{R} 
\end{equation}

Substituting F$_{centrifugal}$ by F$_{gravity}$ 
in the above equation, we can derive an equivalent rotation velocity
V$_{eq}$ due to the mutual gravitational force that will act on material located at a distance
R from the centre in G1. 
If we assume that the gravitational force will be sufficiently strong to affect the material
in the galaxy if V$_{eq} >>$ V $\sim 200$ kms$^{-1}$ for a rotationally supported spiral galaxy
or elliptical galaxy supported by random motions,  then this can be used to estimate
the detectability of the change in angular momentum of the galaxy due to this torque.
V$_{eq}$ estimated for different galaxy masses and gas masses located at a distance of
10 kpc from the centre of the galaxy are tabulated in Table \ref{tab}.
If this distance is increased to 20 kpc, then V$_{eq}$ will increase
by a factor of 1.44 and if the distance is decreased to 5 kpc, then it will
decrease by the same factor. 

\begin{equation}
V_{eq} = \sqrt{\frac{F_{gravity} R}{m}}
\end{equation}

The gravitational torque acting on material at R in G1 due to F$_{gravity}$  is defined as

\begin{equation}
\tau = R~F_{gravity}~sin\theta
\end{equation}

where $\theta$ is the angle between R and F$_{gravity}$ as shown in 
the schematic in Figure \ref{fig2}.  Thus, as long as
$\theta$ is non-zero, the torque due to the gravitational force will
act on the interacting galaxies and modify their angular momentum. 
As long as the gravitational force is sufficient to affect a galaxy, the
torque will be stronger at larger distances from the centre of the galaxy. 

We can also use the arrangement in Figure \ref{fig2} to obtain some 
insights into formation of counter-rotating and ring galaxies. 
Keeping in mind that the torque is perpendicular to the plane 
defined by F$_{gravity}$ and R, we can immediately note that if this
plane is close to parallel to the disk of the galaxy
then the torque in this galaxy will be along or opposite to the angular velocity
vector.  Thus, this torque is likely to modify the angular velocity of the galaxy
resulting in kinematic decoupling and counter-rotation in galaxies.
This is likely to be accompanied by a precession in the gas spin axis (see Figure \ref{fig1}).  
On the other hand, if the plane defined by F$_{gravity}$ and R is perpendicular to
the disk of the galaxy then the torque will be directed perpendicular to the 
angular velocity axis and almost parallel to the disk.  It could be directed inwards or outwards
depending on whether angular momentum is decreasing or increasing.  Thus this torque 
is more likely to modify the distribution of gas ie moment of inertia of the galaxy
and result in ring galaxies or nuclear rings.

\begin{table}
\caption{Equivalent rotation velocity  (V$_{eq}$) of gas of different
masses m at a distance R of 10 kpc from the centre in one of the galaxies
due to gravitational force between the two galaxies.   }
\begin{tabular}{c|c|c|c|c|c|c}
\hline
r kpc  & &   \multicolumn{4}{c|}{V$_{eq}$ kms$^{-1}$} \\
& &   \multicolumn{2}{c|}{M$_1 = 10^{11}$ M$_{\odot}$} & \multicolumn{2}{c|}{M$_1 = 10^{11}$ M$_{\odot}$} &  \\
& &   \multicolumn{2}{c|}{M$_2 = 10^{11}$ M$_{\odot}$} & \multicolumn{2}{c|}{M$_{2}=10^9$ M$_{\odot}$} &  \\
\hline   
& m $\rightarrow$ &  $10^5$ M$_\odot$ & $10^9$ M$_\odot$ &  $10^5$ M$_\odot$ & $10^9$ M$_\odot$  \\
\hline
50  & & 42130 & 421 & 4213  & 42 \\
100 & & 21070  & 210 & 2107 & 21\\
500  & & 4240  & 42 & 424  & 4 \\
1000  & & 2120  & 21& 212  & 2 \\
2000  & & 1025  & 10 & 102 & 1\\
5000  & & 425  & 4 & 43  & 0.4 \\
\hline
\end{tabular}
\label{tab}
\end{table}

As shown in Table \ref{tab} - galaxies of mass $10^{11}$ M$_\odot$ can influence gas masses 
$\sim 10^9$ M$_\odot$ upto a separation of 100 kpc.  Since typical gas mass
of a spiral galaxy is a few times
10$^9$ M$_\odot$ - this indicates the entire gas component of a galaxy 
can be torqued by the gravitational force
when two massive galaxies lie within 100 kpc.  The table further shows that masses of $10^5-10^6$
M$_\odot$ can be torqued even upto separations of 2 Mpc.  Thus, the effect of
a torque due to mutual gravity has high potential to govern the dynamical evolution of galaxies. 
It is remarkable that this simple physical treatment is borne out by observational results 
as discussed in the next sections.

\subsection{A demonstrator: the earth-moon system}
Torques due to the force of gravity are ubiquitous in the
entire solar system and are responsible for tidal locking of the natural satellites
to the planets.  That most planets and their satellites show prograde motion with respect
to the sun but a few (e.g. Venus, Uranus) show retrograde motion indicates that the overall effect is complex.
We confine this discussion to the important evolutionary role played by torques due to gravity in the
earth-moon system.  Although no mass transfer due to gravitational forces has happened in the earth-moon
system, the torque has led to extensive angular momentum exchange in the system.

It has been shown that the torque in the earth-moon
system has caused the moon to gain orbital angular momentum from the earth's spin
angular momentum causing it to move
to a farther orbit.  From Equation \ref{eqn1}, this can be understood as the earth
losing spin angular momentum (decrease in angular velocity) which the moon
is gaining by increasing its separation to the earth (increase in moment of inertia).
No change in the orbital angular velocity of the moon is expected since it is already tidally locked
to the earth.  It is estimated that the earth's rotation is slowing down at a rate of about 1.7 ms per century
\citep{1995RSPTA.351..165S}.  The rotation period of the earth has slowed down since 
its formation from 6h to 24h!

\section{Counter-rotating galaxies: torques}

After having put forward the torque origin scenario to explain counter-rotation in galaxies, we 
tested our hypothesis in the following way.  If the torque due to a closeby
galaxy G$_2$ was responsible for inducing counter-rotation in the gas of galaxy G$_1$, 
then  (1) the counter-rotating galaxies should have a closeby neighbour.
For example, a galaxy moving with a velocity of
200 kms$^{-1}$ would have traversed a distance of 1 Mpc in 6 billion years and would still
be in the neighbourhood.  (2) Since all the counter-rotating galaxies
already have a rotational component, the closeby companions are rotating galaxies.
(3) The companion galaxy shows some signature of the angular momentum exchange.
(4) Since torques will precede tidal stripping, the stellar ages in the counter-rotating
disk should be older than in the tidal features external to the galaxy such as tails and bridges.

\begin{figure}
\begin{center}
 \includegraphics[width=5.5cm]{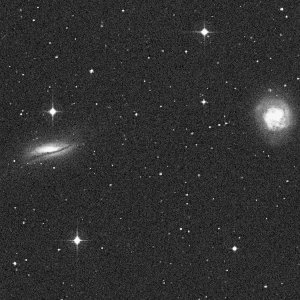}
 \includegraphics[width=5.5cm]{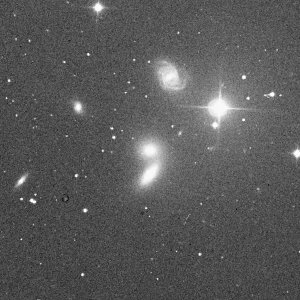}
 \includegraphics[width=5.5cm]{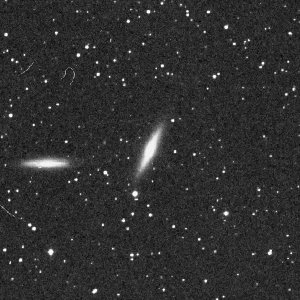}
 \caption{DSS optical images of the counter-rotating galaxies  
NGC 5719 (nearby galaxy: NGC 5713), NGC 5354/NGC 5353, NGC 7332/NGC 7339 of size 15' are shown. 
The bright object northwest of the galaxy in the middle panel with a diffraction pattern is a star.}
 \label{fig3}
\end{center}
\end{figure}

In Figure \ref{fig3}, we show the DSS optical images of the fields of three counter-rotating 
galaxies - NGC 5719,  NGC 5354 and  NGC 7332.  While both NGC 5719 and NGC 7332 show
a closeby companion, several galaxies lie close to NGC 5354.

We examined data on counter-rotating galaxies available in literature or online databases 
- the largest list of 40 counter-rotating galaxies being in the comprehensive review paper by 
\citet{1996ASPC...91..429G}.  We searched recent literature and included galaxies which have since been
found to show counter-rotation or excluded galaxies which have since been found
to not show counter-rotation.  This resulted in a sample of 57 counter-rotating galaxies presented in
Table \ref{tab1}.  We searched for companions in the catalogue of isolated pairs of galaxies
in the northern hemisphere by \citet{1972SoSAO...7....1K}  and in the southern hemisphere by 
\citet{1995ApL&C..30....1R} and for group membership in the LGG catalogue \citep{1993A&AS..100...47G}
and LCDE catalogue \citep{2007ApJ...655..790C}.  
We also did a visual examination of the environment using optical images 
of the Digitised Sky Survey (DSS) downloaded from Skyview. 
We then used the NASA-IPAC Extragalactic Database (NED) to search for the nearest companion 
to the counter-rotating galaxy which had a NGC, UGC or ESO identification.
We retained the default NED search within a velocity range of $\pm 500$ kms$^{-1}$
and a radial distance of 750 kpc.  The heliocentric velocities, distance to the galaxy and separation to
the companion galaxy were recorded from NED.   
The morphological types, position angle, inclination, rotation velocity,
HI mass and the heliocentric velocity determined
using radio data which we assume indicates the velocity of the gas and the heliocentric
velocity determined using optical data which we assume indicates the velocity of the stars
were noted from Hyperleda \citep{2014A&A...570A..13M}.  
All the galaxies are listed in Table \ref{tab1}. 
We note that the list might not be complete since literature is vast. 

\begin{table}
\caption{The counter-rotating galaxies (first entry) and the nearest galaxy
in sky plane from NED (second entry).  Entries in
columns 3,4,5,6 are from HyperLeda and the estimates in
columns 4 and 7 are done using the luminosity distance and angular separation 
given by NED.  Column 5 ($\Delta$V$_{hel}$) lists the difference between the 
heliocentric velocity of the galaxy measured using optical tracers and radio tracers
listed by Hyperleda and dash (-) indicates one of the velocities is not available. }
\begin{tabular}{l|l|c|c|c|c|c}
\hline
No & Galaxy & type & M$_{HI}$ & $\Delta V_{hel}$ & V$_{rot}$ &  d \\ 
     &  &  & $\times10^8$  & \multicolumn{2}{c}{kms$^{-1}$}  & kpc\\
     &  &  &  M$_{\odot}$ & &  & \\
\hline
1 & NGC 128 & S0 & - & 191$^1$  & 141.7 & 13 \\
1 & NGC 127 & S0 & - & 44$^1$  & - &  \\
2 & NGC 5354& S0 & 110 & -176  & 122.5 & 13 \\ 
2 & NGC 5353& S0 & 52 & 77  & 135.7 & \\
3&  NGC 7332 & S0  & 0.32 & -$^2$ & 121.3 & 17.5 \\ 
3&  NGC 7339 & SABb & 2.4 & 14$^2$  & 163.8 & \\ 
4& NGC 1596& S0 & 26 & 64$^3$ & 98.3 & 18 \\ 
4& NGC 1602& SBm & 34 & -127$^3$  & 43.9 &  \\
5 & NGC 770 & E & 21 & -111 & 219.7  & 32 \\
5 & NGC 772 & Sb & 160 & -26 & 260.7 & \\
6 & NGC 4365 & E & - &  - & - & 32 \\ 
6 & NGC 4366 & E & -  & - & - &  \\
7 & UGC 7425 & SABc & - & - & - & 33 \\ 
7 & IC 783 & SB0-a & - & -& - & \\
8 & NGC 4424 & SBa & 0.85 & -7 & 30.5  & 34 \\  
8 & NGC 4417& SB0 & - & - & -  & \\
9 & NGC 4477 & SB0 & - & - & - & 35 \\ 
9 & NGC 4479 & SB0 & - & - & -  & \\
10 &  NGC 3608 & E & 0.29 & -111 & 85.1 & 37 \\
10 &  NGC 3607 & E-S0 & - & -  & - & \\
11 & IC 1459& E & 3.3 & 28 & 36.4  & 40 \\
11 & IC 5264& Sab & 37 & 218 & 230.8  & \\
12 & IC 2006 & ER & 7.2 & 27 & 150.3 & 40 \\ 
12 & ESO359-G005 & I & - & -4 & -  & \\
13 & NGC 3626 & S0-a & 17 & -12  & 190 & 42 \\
13 & UGC 6341 & Sd & 27  & -142 & 43.8 & \\ 
14& NGC 5898& E  & - & - & - & 49 \\ 
14& ESO 514-G003 & E & - & - & - & \\
15& NGC 1216 &  S0-a & - & - & 70.6 & 52 \\
15& NGC 1215 &  SabB & 94 & 72  & - & \\
16& NGC 7097 & E & - & 71 & - &   55 \\ 
16& NGC 7097A & E & - & - & - &  \\
17& NGC 5173 & E & 16 & 7 & 181.4 & 57 \\ 
17& NGC 5169 & SBbc  & - & -15 & 155.8  & \\
18& NGC 936 & SB0-a & 1.4 & 13 & 340.8  & 60 \\ 
18& NGC 941 & SABc & 6.8 & 10 & 89.8 & \\
19& NGC 7331 & Sbc & 15  & 2 & 252.4 & 60 \\ 
19& NGC 7320 &  Scd & - & -13 & 81.8  & \\
20 & NGC 4550 & SB0 & 5 & -54 & 157  & 64 \\  
20 & NGC 4552 & E & - & - & -  & \\
21& IC 4889& E & 83 & 49  & 50.3 & 75 \\
21& IC 4888& E-S0 & - & -  & 50 & \\
22& NGC 4379 & E-S0 & - & - & -  & 80 \\
22& IC 3313 & E & - & - & - & \\
23& NGC 3528 & S0 & - & - & - &  82 \\
23& NGC 3529 & SBb & - & - &   - & \\ 
24&  NGC 5719 & SABa &  120 & 7 & 187.1 & 90 \\  
24&  NGC 5713 & SABb & 67 & -65 & 107.9  & \\
25& NGC 1700  & E & - & - & - & 102 \\
25& NGC 1699  & Sb & - & - & - &  \\
26& NGC 6684& SB0 & - & - & 129.9 & 102 \\
26& NGC 6684A& IB & 8.6 & 1 & 27.1  & \\
27& NGC 4138 & S0-aR & 8.1 & 16 & 150.4  & 120 \\  
27& UGC 7146& I & -  & -  & - & \\
28& NGC 3593 & S0-a  & 4 & -3  & 107 & 150 \\ 
28& IC 2684 & S & - & -12  & 18.6 & \\
\hline
\end{tabular}
\label{tab1}
\end{table}

\begin{table}
\begin{tabular}{l|l|c|c|c|c|c}
\hline
29& NGC 4672 & Sa  & 250 & -2 & 173.8 & 150 \\ 
29& NGC 4677 & SB0-a & -  & - & - & \\
30& NGC 252 & S0-aR & 32 & 16 & 339.1  & 155 \\ 
30& NGC 260 & Sc & - & -8 & 176.6  & \\
31& NGC 5322 & E & - & - & -  & 160 \\
31& UGC 8716 & G & - & - & 48.4  & \\
32& NGC 4826 & Sab & 8.3 & 15 & 152.2  & 170 \\
32& UGC 8011 & I  & 9.8 & -8 & 75.5   & \\
33& NGC 3351 & SBb & 21  & -1 & 149.8 & 185 \\ 
33& NGC 3368 & SABa & - & -15 & 201.6  & \\
34& UGC 6570 & S0-a & 2.4 & -8 & 53.3   & 214 \\ 
34& UGC 6603 & SBc & - & 7 & 79.5  & \\ 
35& NGC 2768 & E  & 3.1 & - &  64  & 235 \\
35& NGC 2742 &Sc & - & 15  & 153.6 & \\
36 & NGC 7079 & SB0 & - & - & 229.1 & 253 \\
36 & ESO287-G037& SBm & 30 & 12 & 67.2 & \\ 
37 & NGC 1574 & E-SB0 & - & - & -  & 290 \\
37 & IC2058 & Scd  & - & 10 & 89.4  & \\
38 & NGC 3941 & SB0 & 3.1 & -15 & 117.4 & 300 \\
38 & NGC 3930 & SABc & - & -4 & 96.5  & \\
39& NGC 3032 & SB0R & 2 & 16 & 153.9 & 324\\
39& NGC 3026 & IB & 30 & 11 & 89.7 & \\
40& NGC 2983 & SB0-a & - & - & 168.2 & 345 \\ 
40& ESO566-G002 & SB0-a  & - & - & -  & \\
41& NGC 3011 & S0R & 2.2 & 12 & 120.8  & 351 \\
41& UGC 5287 & SBc  & 4 & 17 &  97.8  & \\
42 & NGC 4546 & E-SB0 & 2.8 & -21  & 214.4 & 399 \\
42 & UGCA 286 & SBm & -  & 0 & 53.3 &  \\
43& NGC 7377 & SB0-a  & - & - & -  & 410 \\ 
43& ESO603-G015 & S0-a  & - & -  & - & \\
44& IC 456 & SB0  & - & - & -  & 410 \\ 
44& ESO427-G029 & S0 & - & - & -  & \\
45& NGC 4191& S0 & 6  & 24 & 223.3  & 517 \\  
45& NGC 4224 & Sa & - & 12  & 256.7 & \\
46& NGC 4698 & Sab & 18 & 9 & 201.6  & 553 \\ 
46& IC 3767 & E & -   & - & -  & \\
47& NGC 3203 & S0-a & - & - & - & 583 \\ 
47& NGC 3208 & SBc & - & -886$^4$ & 148.1  & \\
48& NGC 497 & SBbcR & 130 & 10 & 278  & 590 \\ 
48& UGC 928 & S0 & - & -  & -  & \\
49& NGC 7796 & E & - & 38  & - & 590 \\ 
49& ESO149-G009 & SBd & - & - & -  & \\
50&  NGC718 &  SaB & 0.2 & -18 & 155.4  & 595 \\ 
50&  UGC1240 & Sd & 8.8 & 10 & 54.3 & \\
51& NGC 4513 & S0  & - &  - & -  & 610 \\ 
51& NGC 4332 & SBa & -  & 7 & 185.8 &  \\
\hline
52 & NGC 3835 & Sab & 12 & 11  & 188.8 & - \\
53 & NGC 6701 & SBaR & 26 & -11 & 242 & - \\
54 & NGC 7007 & E-S0 & -  & - & - & - \\
55 & NGC 7218 & SBc & 15 & -3 & 135.7 & - \\
56 & NGC 7252 & SB0R & 12 & 62  & 165.5 & - \\
57 & NGC 7360 & E & 71 & 15 & 156.9 & - \\
\hline
\end{tabular}

$^1$ The radio HI V$_{hel}$ of 4050 kms$^{-1}$ used for both galaxies taken
from \citet{2012MNRAS.422.1083C}.   \\
$^2$ No HI detected from NGC 7332 and the radio HI V$_{hel}$ of 1300 kms$^{-1}$ for NGC 7339 
taken from \citet{2012MNRAS.422.1083C}.    \\
$^3$ The radio HI V$_{hel}$ of 1621 kms$^{-1}$ for NGC 1621 and of 1584 kms$^{-1}$ for NGC 1596
taken from \citet{2006MNRAS.370.1565C}.    \\
$^4$ Needs to be confirmed. 
\end{table}

\subsection{Results}
The main results of our study of the 57 counter-rotating galaxies can be listed to be:
\begin{enumerate}
\item 45 galaxies have a LGG or LDCE group membership and three of these are also listed in the isolated
pairs of galaxies.  12 galaxies are not listed
as members of a group and NED could not find a neighbouring galaxy with a NGC, UGC or
ESO listing for five of these listed at the bottom in Table \ref{tab1}.
Moreover no nearby galaxy could be found for one of the galaxies (NGC 7252) which has a LGG listing.   
NED could find a companion galaxy for 51 galaxies.  We list all the 57 galaxies in Table \ref{tab1}. 
In some cases, there were several galaxies close to the counter-rotating galaxy
in which case we have included the closest galaxy listed by NED which had either a NGC, UGC or ESO 
listing. 
\item The galaxies are located at distances between 6 and 110 Mpc.  The separation between
the two nearby galaxies in the sky plane ranges from 13 to 610 kpc (column 7 in Table \ref{tab1}).  
The difference in the heliocentric velocities of the two nearby galaxies as listed in NED
varies from 1 to 496 kms$^{-1}$ and the rotation velocity
(column 6 in Table \ref{tab1}), where available, varies from 18 to 340 kms$^{-1}$.   
\item The morphology of the 51 counter-rotating galaxies with a companion galaxy are as follows: 
15 are lenticulars (S0),  13 are ellipticals, 11 are spirals (Sa to Sc), 9 are S0-a,  
and 3 are E-S0.  17 galaxies host a bar.  
\item The morphology of the 51 companion galaxies are: 28 galaxies are spirals 
(Sa to Sd), 6 are ellipticals, 6 are S0s, 5 are irregulars, 3 are S0-a, 2 are E-S0 
and the morphology of one is not listed by Hyperleda. 
Thus, a large fraction of companion galaxies are rotating disk galaxies. 
\item In column 4, the HI mass is estimated from the parameter 'm21c' listed by
Hyperleda which quantifies the HI mass corrected for self-absorption.  36/57
counter-rotating galaxies and 15/57 companion galaxies have HI detections reported in Hyperleda.  
However we are aware that if the measurement is made with a large observing beam then the HI
would be corrupted by nearby galaxies or inter-galaxy medium and should be confirmed.
The HI masses range from $2 \times 10^7 - 2.5 \times 10^{10}$ M$_\odot$. However this
does not include the ionized gas which is detected in several early type galaxies.
\item Since the position angle measures the major axis orientation, we use the difference 
in the position angles of the two galaxies as a measure of the angle
between the rotation axes of the two galaxies.  For the available data on 41 pairs,
we find that the difference for 
28 pairs is between 40$^{\circ}$ and 119$^{\circ}$ and for
13 pairs is between 120$^{\circ}$ and 39$^{\circ}$. 
\item Hyperleda lists the heliocentric velocity of several galaxies
measured using optical tracers and radio tracers.  This, we note, can be understood as
the heliocentric velocities of the stellar disk and the gas disk which in the quiescent case 
should be similar. 
We find that these optical and radio heliocentric velocities show
a difference (see column 5 in Table \ref{tab1}) 
greater than 50 kms$^{-1}$ for several galaxies -seven counter-rotating
and six companions, when both
are available.  
One of the galaxies NGC 7252 for which no companion galaxy was found within 750 kpc by NED
also shows a difference of 62 kms$^{-1}$. 
At least one of the galaxies in the counter-rotating pair shows such a large offset 
($\ge 50 $ kms$^{-1}$) if the separation between them is less than 30 kpc and 
several galaxies whose separation is less than 100 kpc show such an offset.  No such
offset is observed for galaxies separated by more than 100 kpc.
While several HI measurements might be uncertain due to a large observation beam, we do
believe that this result is significant. 
A targetted study using interferometric HI observations
should be undertaken such as in \citet{2012MNRAS.422.1083C}.  We have modified the entries in
column 5 of Table \ref{tab1} for NGC 128/NGC 127, NGC 7332/NGC 7339 and NGC 1596/NGC 1602 from
the HI results of \citet{2012MNRAS.422.1083C} and \citet{2006MNRAS.370.1565C}.   
No HI is detected from  NGC 7332 whereas marginal detection of HI is reported from NGC 128
and NGC 1596. 

\end{enumerate}

Thus from the existing sample of counter-rotating galaxies, we note that 90\% 
have a rotating companion galaxy within a projected distance of 750 kpc which as noted above
strongly favours the torque theory.  In particular, 25 galaxies have a companion
within 100 kpc and many of them show an offset 
between the radio and optical velocities.  This difference between the radio
and optical velocities could be indicative of a
change in the orbital angular momentum of the galaxies and further
observations to confirm this need to be undertaken.

\subsection{Case studies - NGC 5719/NGC 5713 and NGC 4550}
The top panel in Figure \ref{fig3} shows the DSS optical image of the young pair of galaxies,
NGC 5719 and NGC 5713.  Large quantities of hydrogen
has been detected from the galaxies 
and the gas has been found to be counter-rotating with
respect to the main stellar disk in NGC 5719 \citep{2007A&A...463..883V}.  
HI is also detected from between the galaxies \citep{2007A&A...463..883V}. 
The old stellar disk in NGC 5719 is in retrograde rotation with NGC 5713 while the
counter-rotating gas disk and 20\% of stars in NGC 5719 are in prograde rotation with NGC 5713. 
We suggest that the torque acting on NGC 5719 due to NGC 5713 has modified the spin
angular momentum of NGC 5719 leading to a change in direction of rotation of the gas disk. 
Moreover the HI major axis of NGC 5719 (PA=$121^{\circ}$) makes an angle of $25^{\circ}$ with 
the optical disk (PA=$96^{\circ}$) \citep{2007A&A...463..883V}.  
As an effect of the counter-torque on NGC 5713,  we find that there is an offset  of
65 kms$^{-1}$ in it optical and radio heliocentric velocities listed by Hyperleda. 
Thus, we infer that the torques have led to counter-rotation in NGC 5719 and change in
its gas spin axis ie
change in its spin angular momentum whereas a torque on NGC 5713 has led to
a decoupling in the systemic motions of the gas and stellar disks. 

The age of stars in the counter-rotating disk of NGC 5719 is
between 300 Myr and 700 Myr \citep{2005ApJ...619L..91N,2011MNRAS.412L.113C}.
On the other hand the age of stars formed in the tidal features external
to the galaxies is between 2 Myr and 200 Myr \citep{2005ApJ...619L..91N}.  
Thus, the age of the counter-rotating population of stars in NGC 5719 is older than
in the tidal features.  This is what is expected in the torque origin since
there is no external inflow of gas but the galaxy's own gas is torqued to counter-rotation.  
The star formation in the torqued material should precede the tidal 
stripping of material from the galaxies as witnessed in this pair. 
We note that this is difficult to explain in the existing gas accretion model where  
the stellar population in the external tidal features should be
older than in the counter-rotating disk since the counter-rotating material is 
being accreted into NGC 5719.  Thus the tidal features external to NGC 5719 would
have formed well before the galaxy accreted this material. 
We, thus, find that strong support for the gravitational torque leading to counter-rotation 
comes from the ages of the stellar populations.

Having discussed the above, we go on to check if this system satisfies the four criteria
that we had put forward at the start of this section: (1) NGC 5719 has a neighbour
(2) NGC 5713 is a rotating companion galaxy  (3) NGC 5713 shows a decoupling between the
systemic motion of the gas and optical disks which
can be taken as a signature of the torque on this galaxy  (4) The stellar ages in the
tidal features external to NGC 5719 are younger than the counter-rotating stars in NGC 5719.  
Thus this pair of galaxies satisfies all the four criteria that we had put forward for a 
torque origin.  

For comparison, we check if the pair satisfies the gas accretion/merger model
especially observables (1) and (3) listed in section 1.  The morphology of the counter-rotating
galaxy NGC 5719 is undisturbed and is gas-rich
\citep[M$_{HI}=7.2\times10^9$M$_\odot$;][]{2007A&A...463..883V}.  
In the gas accretion model, the counter-rotating gas should have been accreted from NGC 5713 but 
we note that NGC 5713 is also gas rich \citep[M$_{HI}=6.6\times10^9$M$_\odot$;][]{2007A&A...463..883V}
and does not seem to have lost much of its gas.  
In fact the HI in the pair strongly supports the scenario in which both the galaxies still 
possess their own gas and stellar disks and the gas has started
being tidally stripped from both galaxies some 200 Myrs ago.   
As pointed out by \citet{2007A&A...463..883V} there is a 74 kpc long tail of HI which
connects the two galaxies along the optical major axis of NGC 5719 which could
be indicative of gas flow from one to the other or to the inter-galactic medium.
We believe that behaviour similar to that exhibited by NGC 5719 and NGC 5713 
will be present in other cases of counter-rotation where both the galaxies are gas rich and
needs to be investigated - if found to be commonly true, it would unequivocally support
the torque origin for observed counter-rotation in galaxies. 

\citet{2007A&A...463..883V} suggest that NGC 5719 
will eventually evolve into a galaxy like NGC 4550 in which
half the stars have been found to be counter-rotating \citep{1992ApJ...394L...9R}
with the counter-rotating stars younger at 2.5 Gyr and the
main stellar disk being older at 11 Gyr \citep{2013MNRAS.428.1296J}.  
Retrograde motion in the ionized gas and half the stars in NGC 4550 has been explained
by gas accretion \citep{1992ApJ...394L...9R}.  
We suggest that the origin of the counter-rotating matter would have been 
in a gravitational torque and similar to NGC 5719 - however
the interaction would have happened more than 2.5 Gyr ago.  We examine NGC 4550
in terms of observables.  A companion galaxy NGC 4552 is found
to be located at a projected distance of 64 kpc from NGC 4550.  A difference of 54 kms$^{-1}$
between the optical and radio heliocentric velocities is found for NGC 4550.
The companion galaxy NGC 4552 is an elliptical which shows rotation in the ionized gas 
\citep{2009APS..OSF.P1005K}. 
No HI \citep{2012MNRAS.422.1835S} or molecular gas \citep{2011MNRAS.414..940Y} have been detected 
from either NGC 4550 or NGC 4552.  Thus the torque acting on NGC 4550 has resulted in
a counter-rotating component and a kinematic offset between its gas and stellar disks. No
signature on the companion galaxy could be found but 
more data on NGC 4552
are required to search for signatures of this interaction on its morphology or kinematics. 
It is important to devise smart observations which are sensitive to
signatures which can distinguish between the different formation scenarios. 

We note that the order-of-magnitudes estimates listed in Table \ref{tab} 
successfully explain observations of NGC 5719 and NGC 5713.  Both are massive galaxies,
are separated by 90 kpc and the entire gas disk in NGC 5719 is counter-rotating.  
From Table \ref{tab}, we infer that the torque exerted by the gravitational force between two
massive galaxies ($\sim 10^{11}$ M$_\odot$) if separation $<$ 100 kpc can 
affect the entire gas disk ($\sim 10^9$ M$_\odot$).  

\section{Ring galaxies: Torques}
We examined other observed phenomena which can be explained by torques mediating angular momentum
exchange. In particular, we search for observables where the angular momentum
exchange is mediated not through changes in $\omega$ but through changes in the moment of inertia of
a galaxy ie in the radial distribution of material (see Equation \ref{eqn1}).  
Existence of such systems would provide independent support to the torque origin.
Interestingly,  we find that the observations of collisional ring galaxies where
the ring is observed in the galactic plane support an origin in 
torques.  In this case, the gas observed along the ring
would have originated in the galaxy itself and the name 'collisional' would be a misnomer.  
However we continue to use it here to distinguish it from the resonance 
rings due to the bar. 
No external gas accretion or collision with a dwarf galaxy are required in the proposed
torque origin.

The collisional ring galaxies have generally been explained either 
due to a head-on collision of a disk galaxy with another
galaxy using the observation that a companion galaxy is generally found along the minor
axis and simulations \citep{1976ApJ...208..650T,1976ApJ...209..382L} or by collision with intergalactic HI
clouds \citep{1974ApJ...194..569F}.  However \citet{1976SvAL....2..204V} had suggested that the
ring galaxies are not generated by collisions but by more quiescent physical phenomena
internal to the ring galaxy.  On the other hand, rings termed as resonance rings 
and found to be of three different types - inner, outer and nuclear \citep[e.g.][]{1995ApJS...96...39B}
are believed to be formed due to orbital resonance with a bar \citep{1996FCPh...17...95B}.  
While examining the ring galaxies shown in \citet{1995ApJS...96...39B}, we note that
some of the galaxies show an outer ring which appears similar to the collisional ring thus cautioning
that while the rings observed encircling the bar in barred galaxies are likely
to be generated by processes associated with the bar, the outer and nuclear rings might 
be a result of the action of gravitational torques due to a neighbouring galaxy.    

In the last section we demonstrated how the angular momentum exchange
due to a torque naturally explains the observed counter-rotation 
and how observations support this origin.
To understand the ring galaxies in terms of action of torques, we again refer to the two terms in
Equation \ref{eqn1} and Figure \ref{fig2}.  To recall, the torque that acts on a galaxy can be
manifest as a change in the angular velocity or a change in the distribution of matter.  
The moment of inertia of the galaxy can change due to redistribution of matter - expansion
or contraction of an annular ring. 
The observed structure of ring galaxies where most of the gas is accumulated along a ring outside
the optical galaxy and lower column density HI is measured between the ring and the
galaxy can be explained by the gas moving out to larger radial distances under the influence of
a torque.  If we rearrange Figure \ref{fig2} so that the companion galaxy is 
along the minor axes of G1 then F$_{gravity}$ will be perpendicular to the
disk of G1.  The plane made by F$_{gravity}$ and R will also be perpendicular to the disk of
G1 and the torque will be almost parallel to the disk causing matter to radially expand or contract.
Strong support for this origin comes from
the observational result that a companion galaxy is almost always located along the minor axis of
ring galaxies \citep{1976ApJ...208..650T}. 

Interestingly, some of the counter-rotating galaxies also show HI distributed along rings e.g. NGC 4826;
\citep{1991A&A...245....7V} and NGC 3941 \citep{1989A&A...225..317V}.  In 
the case of NGC 4826, the inner HI disk is counter-rotating with respect to the outer HI
disk \citep{1992Natur.360..442B}.  We suggest that there might be more common
cases of counter-rotation and ring galaxies since both are mediated by a torque and
the angular momentum change can be in both the angular velocity and radial distribution of the gas disk.
However we note that that both phenomenon in the same galaxy would be difficult to
explain in the gas accretion model.  We could not find any of the galaxies listed
in Table \ref{tab1} in the catalogue of collisional ring galaxies \citep{2009ApJS..181..572M}.
We reiterate that the counter-rotating components and ring
structures can all be consistently explained by invoking torques acting on the galaxy.

Another important common property of ring galaxies and galaxies which show counter-rotation is 
the presence of a nearby companion.  
We claim that this is a stronger proof of gravitational torques being responsible for the ring
galaxies than a collision encounter since in the latter the colliding galaxy can merge, 
disintegrate or not be detectable.  Hence we suggest that a companion galaxy is not a pre-requisite
in the collision scenario.  In fact, the presence of a companion for all the ring galaxies itself
makes the collision scenario suspect since at least a few should have lost their identity
in the collision.  Even if we assume that all the intruder galaxies are intact after
the head-on collision, these intruders should certainly lose their gas to the ring galaxy.
Thus, an important observational test of the collision model would be that all
the galaxies if they are intruders should be gas-poor.  While this does not rule out the torque origin,
it lends observational support to the collision model although the question still remains
how all the intruder galaxies have managed to remain intact after the collision.  
Since \citet{2009ApJS..181..572M} have listed the
companion galaxy alongwith the ring galaxy, this test should be easily doable. 
We could not locate a study of this kind in literature. 

In the torque origin, gas of the parent galaxy
forms the ring and there is no need for a collision.  Strong support for this origin is also evident
in the observation that many of the ring galaxies are gas-poor S0s or ellipticals
and it is likely that the gas from these galaxies has been prematurely moved out 
by the torques (see Equation \ref{eqn1}), thus accelerating the evolution of the optical
galaxy to a gas-poor state.   Thus, an important formation scenario
of lenticular galaxies might be the action of torques manifest on its gaseous component which pushes
it out in form of rings which are eventually removed from the galaxy.
The rings continue to show expansion and while these might survive longer in small groups or pairs of galaxies,
the rings are likely to be stripped close to cluster centres due to ram pressure stripping
\citep{1972ApJ...176....1G} or fast galaxy encounters ie harassment \citep{1996Natur.379..613M}
leading to the observed morphology-density relation in clusters \citep{1980ApJ...236..351D}. 

We find it interesting that \citet{2010A&A...519A..40A}, from their observational study of ionized gas in 66 
nearby lenticular/elliptical galaxies have
noted that $\sim 50\%$ of these show some peculiarities in kinematics such as counter-rotation or
in features or decoupled gas distributions with respect to the stars.  
We suggest that many of these are signatures of the galaxy being subjected to a
gravitational torque.
In an independent study of similar galaxy types, \citet{2011MNRAS.417..882D}
find that 36\% of the fast rotating sample galaxies, but none a massive fast rotator, 
show misalignment of kinematic axis of the stellar and gas disks (see Figure \ref{fig1}). 
These results naturally follow if the misalignment is caused by a mutual gravity torque leading
to a change in the angular momentum of the galaxy.
The observable effect of this torque will be 
small on the massive (large M) fast rotating (large rotation velocity) 
galaxies as can be inferred from Eqn 4 and listed in Table \ref{tab} unless the gas mass
is small.

We discuss the torque origin of the ring in the Cartwheel ring galaxy below. 

\subsection{A case study: the Cartwheel galaxy}
\begin{figure}
\begin{center}
 \includegraphics[width=6.5cm]{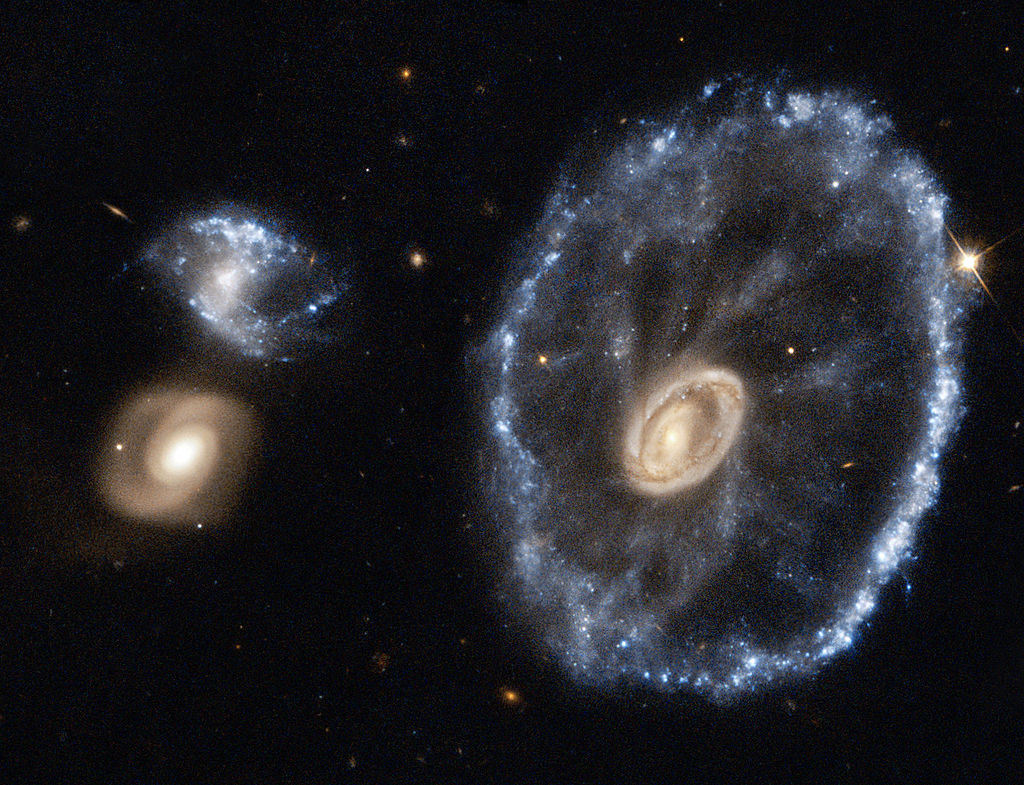}
 \caption{ HST image of  the Cartwheel galaxy and two other member galaxies PGC 2249, PGC 2252 of the
group SCG 0035-3357 downloaded from the internet.  
The fourth member PGC 3170815 which is to the north is not shown. }
 \label{fig4}
\end{center}
\end{figure}

One of the most famous ring galaxies is the Cartwheel galaxy shown in Figure \ref{fig4}.  
The Cartwheel galaxy along with three other galaxies namely, 
PGC 2249, PGC 2252 and PGC 3170815  
form the group of galaxies SCG 0035-3357 \citep{2002AJ....124.2471I} or the Cartwheel group (Figure \ref{fig4}).
Details of the members are listed in Table \ref{tab2}. 
HI has been detected from all the galaxies \citep{1996ApJ...467..241H}.
Star formation has been triggered in the gas in the ring and
H$\alpha$ emission has been detected from
the outer and inner rings in the Cartwheel galaxy and from PGC 002249 \citep{1998A&A...330..881A}.  
The Cartwheel galaxy structure has been explained as being a result of collision with 
another galaxy \citep[e.g.][]{1993ApJ...411..108S,2001Ap&SS.276.1141H}.  
We examine a torque origin for the observed gas properties of the galaxies in the group and 
the ring.

Cartwheel galaxy shows the presence of a ring surrouding the optical galaxy and spoke-like structures
which seem to connect the ring to the galaxy.   Most of the HI in this galaxy is
found in the ring encircling the galaxy and there is low column density HI between the optical
galaxy and the ring \citep{1996ApJ...467..241H}.  In particular we note: (1)  The HI in the ring shows a
combination of rotation (V$_{rot}=291$ kms$^{-1}$) and expansion (V$_{exp}=53$ kms$^{-1}$) and 
a systemic velocity of 9089 kms$^{-1}$ \citep{1996ApJ...467..241H}.  
The H$\alpha$ gives the following V$_{rot}$=217 kms$^{-1}$, V$_{exp}=13-30$ kms$^{-1}$ and a systemic
velocity of 9050 kms$^{-1}$ \citep{1998A&A...330..881A}.
In the torque origin, an increase in the angular momentum can be inferred from the gas redistribution 
to larger radial distances.  Thus, the gas in the ring should display both differential rotation and expansion
as is observed.  (2) In Table \ref{tab2}, the difference in the optical and radio heliocentric
velocities are listed in column 5.  While all members show an offset $>20$ kms$^{-1}$, PGC 3170815
shows a large offset of 120 kms$^{-1}$. 
This can indicate the effect on this companion of a torque by Cartwheel galaxy.   
Moreover the HI disk of PGC 3170815 is displaced to the
north compared to the optical disk but no distortion in its velocity field is observed 
\citep{1996ApJ...467..241H}.  Both these point to a change in its 
orbital angular momentum.  In fact by combining these two observables, it can be inferred that
the HI disk is receding to the north with a velocity of 120 kms$^{-1}$ and shows a detectable 
displacement. 
(3) The HI kinematic and photometric axes of PGC 2249 show a difference of 15$^{\circ}$
\citep{1996ApJ...467..241H}.  
\begin{table}
\caption{The four member galaxies of the poor young group Holmberg 124.
Entries in columns 3 and 6 are from HyperLeda and in columns 4
is from \citet{1996ApJ...467..241H} for Cartwheel galaxies and from \citet{2005A&A...435..483K}
for Holmberg 124 galaxies.  Column 7 is from NED.
For the Cartwheel group, column 5 lists the difference between optical and radio HI velocities taken from
\citet{1996ApJ...467..241H} and for Holmberg 124, between optical and radio HI velocities
from  \citet{2005A&A...435..483K}.}
\begin{tabular}{l|l|c|c|c|c|c}
\hline
No & Galaxy & type & M$_{HI}$ & $\Delta V_{hel}$ & V$_{rot}$ &  d \\
     &  &  & $\times10^8$  & \multicolumn{2}{c}{kms$^{-1}$}  & kpc\\
     &  &  &  M$_{\odot}$ & &  & \\
\hline
\multicolumn{7}{c}{Cartwheel group} \\
1 & Cartwheel & E-S0 & 93  & 39  & 149.9 & -\\
2 & PGC 2249 & SBab & 27 & 28 & - &  36\\
3 & PGC 2252 & SB0-a & <0.1 & -24   & - & 36 \\
4 & PGC 3170815 & Sb & 7 & 120 & - & 118 \\
\hline
\multicolumn{7}{c}{Holmberg 124 group}\\
1 & NGC 2820 & Sc & 66  & 3  & 162.8  & 0 \\
2 & Mrk 108 & S0-a & 0.6 & -117 & - & 13 \\
3 & NGC 2814 & Sb  & 3.4  & 115 & 160.6 & 25 \\
4 & NGC 2805 & SABc & 79$^1$  & 12 & 70.4 & 88 \\
\hline
\end{tabular}

{\footnotesize $^1$ This is taken from Hyperleda due to missing flux in \citet{2005A&A...435..483K}.} 
\label{tab2}
\end{table}

All these observational signatures involving gas in the Cartwheel group of
galaxies are expected observables if the torque is assumed to modify the angular
momentum of the galaxies.   Except for (1), we note that
the other observables in Cartwheel group galaxies cannot be explained by the collision model. 

After having explained the HI observations of the Cartwheel group members using the torque origin,
we examined the general features of ring galaxies listed in Section 1 and find that all the
points can be explained in a torque origin. 

\section{Revisiting Holmberg 124}
In light of the importance of gravitational torques in galaxy evolution and resulting
observables that we have pointed out here, we reexamine the
results on the gas-rich group Holmberg 124 which consists of four late type galaxies - NGC 2820, 
NGC 2805, NGC 2814 and Mrk 108.  Interpretation based on a combined study of the radio continuum,
HI 21cm imaging and optical data from literature was presented in
\citet{2005A&A...435..483K} and \citet{2012BASI...40..515M}.  They interpreted the observables
to be a combination of features due to the physical processes of ram pressure and tidal forces. 
We revisit this young group of galaxies and modify their interpretation.   In particular, we
find that a few features that they attributed to ram pressure-driven effects can be explained 
in a mutual gravity torque origin and constitute a more likely explanation.  
Details for the member galaxies are presented in Table \ref{tab2}.

As seen in Table \ref{tab2} the smaller galaxies NGC 2814 and Mrk 108 show
a large offset between its optical and radio heliocentric velocities.  Additionally as seen in
Figure 7 of \citet{2005A&A...435..483K}, the HI gas of NGC 2814 and Mrk 108 is displaced from the optical
disks.  As discussed in this paper, both these signatures 
are expected due to a change in the orbital angular momentum of the smaller
galaxies 
due to the gravitational torque.  \citet{2005A&A...435..483K} found
it difficult to explain the northern arc of HI clumps observed in NGC 2805.  We suggest that
this arc-like feature or broken ring can be seen to be indicative of the effect
of a torque on NGC 2805 which is pushing HI gas to larger radial distances and in due
course the galaxy might become a ring galaxy.  A comprehensive optical/radio HI
study of gas rich groups of galaxies should be undertaken to statistically check these
observables and their origin and compared to a control sample of binary
and isolated galaxies. 
We are convinced that the torque origin will be able to account for several intriguing observables.

\section{Summary and conclusions}
In this paper, we have presented observational evidence to support a gravitational torque origin 
for the observed phenomena of counter-rotation in galaxies and
formation of ring galaxies.  The galaxy observables which could be a result of torques 
causing a change in its spin or orbital angular momentum are summarised
in Figure \ref{fig1}.  No extra-galactic gas accretion, merger or collisions are required in
this origin.  The torque precedes any tidal stripping of gas
and mediates an angular momentum transfer - spin, orbital or both - between the galaxies. 
We draw inspiration from the earth-moon system where torques have 
successfully mediated exchange of angular momentum between the two bodies 
with no stripping of matter involved. 
We stress that similar physical processes are at work on galaxy scales where more massive
objects are involved and govern the dynamical evolution of galaxies.   
The proposed gravitational torque scenario is schematically shown in Figure \ref{fig2}. 
The summary of the results presented in the paper are:

\begin{enumerate}
\item We suggest that counter-rotation in galaxies and formation of ring galaxies is 
because of a torque due to mutual gravity modifying the spin angular momentum of these galaxies. 
The resulting change in angular momentum is clearly observed in the gas component of
the galaxy, as expected. 

\item 
Counter-rotation in galaxies and ring galaxies are primararily observed
in the gas disk and the main stellar disk of most galaxies appear undisturbed.  
This is expected since the mutual gravity torques will show the first effects on the gas in the galaxies.
Moreover since no external gas/collision/merger
is required in the torque origin, no distortion in the stellar disk is expected. 

\item
Torques due to mutual gravity require the presence of companion galaxies in the immediate environment. 
We collect a list of 57 galaxies from literature which have been documented to show counter-rotation and
find that 51 of these have at least one companion within 750 kpc.  The companion galaxies
are predominantly a rotating spiral galaxy.
All the ring galaxies also have a companion galaxy. 
This observable lends strong support to the torque origin.  

\item
The gravitational torque origin should leave behind a signature on the companion galaxy as summarised
in Figure \ref{fig1}.  We find several observations give evidence of the same:  
(1) an offset between the heliocentric velocities of the stellar and gas disks.
For the counter-rotating galaxy pairs, the offset is $\ge 50$ kms$^{-1}$ 
for many galaxies for a projected sky separation less than 100 kpc.
No significant offset is seen for cases where the separation is more than 100 kpc.  
(2) The gas disk is displaced with respect to the optical disk  (3) Misalignment of the
HI kinematic axis and the photometric axis.  All these observables in the companion
galaxies can be consistently explained
due to a change in the angular momentum of the galaxy induced by the torque. 

\item
In the case of the counter-rotating galaxy NGC 5719 and its companion NGC 5713, we find the following results
in literature:
(1) The age of counter-rotating stellar population $>$ stellar age in the tidal features external
to NGC 5719.  This is as expected in the torque origin and opposite of what would be expected
in a gas accretion model.  (2) The optical disk and HI disk in NGC 5719 are misaligned by $25^{\circ}$. 
(3) The HI gas disk of NGC 5713 has an extra motion of 65 kms$^{-1}$ towards us as compared to its
optical disk.  
All these results are consistently explained if the galaxies are subject to a gravitational torque
exerted by a companion galaxy.

\item
In case of the Cartwheel galaxy which has a gas ring, we find the following from literature:
(1) An outer ring and an inner ring are observed in the Cartwheel galaxy and it has three companion galaxies.
(2) An offset of $15^{\circ}$ is found between the photometric and HI kinematic axes in PGC 2249. 
(3) An offset of 120 kms$^{-}$ is noted between the radio and optical heliocentric velocities of PGC 3170815.
The HI disk of this galaxy is displaced to the north of the optical disk. 
All these results are consistently explained if the galaxies are subject to a gravitational torque
exerted by a companion galaxy.

\item 
We have shown that several galaxy observables can be explained by a torque exerted by
mutual gravity between galaxies changing the angular momentum of the galaxies
which is readily observable in the gas kinematics or distribution.
While this paper discusses the phenomena of counter-rotation and ring galaxies
and the effect of the torque on the companion galaxies, we believe that
gas warps in galaxies (e.g. NGC 4013) and formation of polar ring galaxies
(e.g. NGC 660) can also be due to a change in the angular momentum of
the galaxies due to a gravitational torque.  Even formation of lenticular galaxies can be
explained through the ring galaxy phase. However these need to
be examined in detail. 

\item We end by stressing the importance and widespread influence of mutual gravity torques on the 
dynamical evolution of galaxies. 
A variety of observables in galaxies can be elegantly, simply, consistently and physically
explained in a torque origin.  There appears no need for all these peculiar galaxies to
have had a violent past. 

\end{enumerate}

\section*{Acknowledgements}
I thank the anonymous referee whose comments have helped clarify the paper.
I thank Prasad Subramanian for discussions especially on the amazing physical concept of torque.
I acknowledge generous use of NED (https://ned.ipac.caltech.edu/),
HyperLeda (http://leda.univ-lyon1.fr), ADS abstracts, arXiv, Wikipedia
and Google search engines in this research.

\bibliography{galaxies}

\end{document}